\documentclass[superscriptaddress,notitlepage,onecolumn,amsmath,amssymb,floatfix,nofootinbib]{revtex4-1}
\usepackage{graphicx}
\usepackage{dcolumn}
\usepackage{bm}
\usepackage{hyperref}
\usepackage{rotate}
\usepackage{fancyhdr} 
\usepackage{cancel}
\usepackage{aas_macros,braket}

\def\be{\begin{equation}}
\def\ee{\end{equation}}
\def\bea{\begin{eqnarray}}
\def\eea{\end{eqnarray}}

\def\VEV#1{\left\langle #1 \right\rangle}

\newcommand{\fsky}{f_{\rm sky}}

\newcommand{\bfell}{\boldsymbol{\ell}}

\def\up{\;\raise1.0pt\hbox{$'$}\hskip-6pt\partial\;}
\def\down{\;\overline{\raise1.0pt\hbox{$'$}\hskip-6pt\partial\;}}

\begin{document}

\title{Violation of statistical isotropy and homogeneity in the 21-cm power spectrum}

\author{Maresuke Shiraishi}
\affiliation{%
Kavli Institute for the Physics and Mathematics of the Universe (Kavli IPMU, WPI), UTIAS, The University of Tokyo, Chiba, 277-8583, Japan
}

\author{Julian B. Mu\~noz}
\affiliation{
Department of Physics and Astronomy, Johns Hopkins University, 3400 N.\ Charles St., Baltimore, MD 21218, USA
}

\author{Marc Kamionkowski}
\affiliation{
Department of Physics and Astronomy, Johns Hopkins University, 3400 N.\ Charles St., Baltimore, MD 21218, USA
}

\author{Alvise Raccanelli}
\affiliation{
Department of Physics and Astronomy, Johns Hopkins University, 3400 N.\ Charles St., Baltimore, MD 21218, USA
}

\date{\today}

\begin{abstract}
Most inflationary models predict primordial perturbations to be
statistically isotropic and
homogeneous. Cosmic microwave background (CMB) observations,
however, indicate a possible departure from statistical isotropy
in the form of a dipolar power modulation at large angular scales. 
Alternative models of inflation, beyond the simplest
single-field slow-roll models, can generate a small power
asymmetry, consistent with these observations. Observations of
clustering of quasars show, however, agreement with statistical isotropy
at much smaller angular scales. Here, we propose to use off-diagonal
components of the angular power spectrum of the 21-cm
fluctuations  during the dark ages to test this power
asymmetry.  
We forecast results for the planned SKA radio
array, a future radio array, and the cosmic-variance-limited
case as a theoretical proof of principle.
Our results show that the 21-cm line power spectrum will enable
access to information at very small scales and at different
redshift slices, thus improving upon the current CMB constraints
by $\sim 2$ orders of magnitude for a dipolar asymmetry and by $\sim 1-3$ orders of magnitude for a quadrupolar
asymmetry case. 
\end{abstract} 


\maketitle

\section{Introduction}

One of the key principles of cosmology is the notion of isotropy
and homogeneity---there is no preferred location nor preferred
direction in the Universe.  This, though, is violated by small
primordial perturbations.  Still, the prevailing single-field
slow-roll model for the origin of these perturbations predicts
that isotropy and homogeneity should be preserved in a {\it
statistical} sense. A significant detection
of a deviation from statistical isotropy or homogeneity would falsify some of
the simplest models of inflation, making it necessary to
postulate new physics, such as nonscalar degrees of
freedom.  Moreover, it would open a window into the physics of
the early Universe, shedding light upon the primordial degrees
of freedom responsible for inflation. Departures from
statistical isotropy and homogeneity can take different
forms. We study the two main cases, a dipolar power
asymmetry and a quadrupolar asymmetry.

A detection of dipolar power asymmetry, i.e., a different power spectrum in two opposite poles of the sky, was reported analyzing the off-diagonal $\ell_1 = \ell_2 \pm 1$ components {of angular correlations of the cosmic microwave background (CMB) anisotropies, $\Braket{a_{\ell_1 m_1} a_{\ell_2 m_2}}$,} with the WMAP and {\it Planck} data at large scales \cite{Hansen:2004vq, Gordon:2005ai, Eriksen:2007pc, Gordon:2006ag, Ade:2013nlj,Akrami:2014eta,Ade:2015lrj,Ade:2015sjc,Aiola:2015rqa}, showing an amplitude 3$\sigma$ away from zero. This detection, however, contradicts observations at smaller scales, where 
the distribution of quasars at later times was studied in
Ref.~\cite{Hirata:2009ar} and shows no asymmetry.  This power
asymmetry in the two-dimensional CMB sky is modeled as a spatial
modulation of three-dimensional power.  It thus represents,
strictly speaking, a violation of statistical homogeneity (SH) rather than statistical isotropy.  Given, though, that that SH
violation is manifest as a two-dimensional power asymmetry, we
refer to it here, for simplicity, as a dipolar power asymmetry or dipolar violation of rotational symmetry.

It is then interesting to understand which nonstandard
inflationary models could produce the required departure from SH
as observed in CMB maps, while keeping the quasar bound fixed.  
A variety of models have been investigated in the past, including cosmological defects \cite{Jazayeri:2014nya}, tensor-mode contributions \cite{Namjoo:2014pqa}, a modulated scale-dependent isocurvature contribution to the matter power spectrum or a modulation of the reionization optical depth, gravitational wave amplitude, or scalar spectral index \cite{Dai:2013kfa}. In the last case, one would expect the dipolar asymmetry to cancel at intermediate scales, while being present at very large and very small scales. 
It is interesting to note that some of these scenarios will generate both dipolar power asymmetry and non-Gaussianity~\cite{Schmidt:2012ky,Lyth:2013vha, Abolhasani:2013vaa}.
Some of these models also predict a quadrupolar power asymmetry (see e.g.~\cite{Bartolo:2012sd,Bartolo:2013msa,Shiraishi:2013vja,Shiraishi:2013oqa,Abolhasani:2013zya,Bartolo:2015dga}).

Quadrupolar modulations in the power spectrum of primordial curvature and tensor perturbations can arise in inflationary models where the inflaton field $\phi$ couples to a U(1) vector field $A_\mu$. These modulations are created by the directional dependence of the vacuum expectation value of $A_\mu$. The amplitude and scaling of such modulations are strongly model dependent. In well-known models involving an inflaton-vector interaction, $f(\phi)F^2$, the scale dependence is determined by the time dependence of the coupling function $f(\phi)$ (see e.g.~\cite{Dimastrogiovanni:2010sm,Soda:2012zm,Maleknejad:2012fw,Bartolo:2012sd,Naruko:2014bxa}). In models that include a pseudoscalar field, an axial coupling term like $g(\phi) F\tilde{F}$ can amplify the quadrupolar modulation part of the curvature power spectrum via the enhancement of the mode function of the vector field \cite{Bartolo:2014hwa,Bartolo:2015dga,Abolhasani:2015cve}; a quadrupole can also be generated by an inflating solid or elastic medium \cite{Bartolo:2013msa, Bartolo:2014xfa}. These quadrupolar modulations create CMB angular power spectra in $\ell_1 = \ell_2 \pm 2$, as well as in the diagonal $\ell_1 = \ell_2$ ones \cite{Ackerman:2007nb,Pullen:2007tu,Watanabe:2010bu,Bartolo:2014hwa}. 
Measurements of CMB maps have not detected evidence of such off-diagonal signals; thus, we only have upper limits on the magnitude of the quadrupolar asymmetry \cite{Kim:2013gka, Ade:2015lrj,Ade:2015sjc}. Nonetheless, it is of paramount importance to constrain these quadrupolar modulations, as they would necessarily imply a departure from the standard single-field inflation model.

In this paper, we set out to constrain rotational invariance of the Universe using a different probe, the power spectrum of 21-cm fluctuations during the dark ages.\footnote{See e.g., Refs.~\cite{Hirata:2009ar,Pullen:2010zy,Baghram:2013lxa,Alonso:2014xca,Pitrou:2015iya,Pereira:2015jya,Hassani:2015zat,Zibin:2015ccn,Bengaly:2015xkw,Shiraishi:2015lma} for studies on other observables of statistical anisotropy or inhomogeneity.} The dark ages are the cosmic era following primordial recombination and ending when the first luminous objects are formed.
CMB photons with a rest wavelength of 21 cm during this era can get resonantly absorbed by neutral hydrogen gas. This process imprints the fluctuations present in the hydrogen density into the observed temperature anisotropies \cite{Loeb:2003ya,Bharadwaj:2004nr}.
In this case, one can then reach angular scales far beyond those available to CMB measurements, as the theoretical limit is given by the Jeans scale, at $\ell\sim 10^{6-7}$ \cite{Lewis:2007kz}.
It has recently been proposed to use this cosmological observable to study a variety of physical phenomena, such as primordial non-Gaussianities \cite{Cooray:1999kg,Cooray:2004kt,Cooray:2008eb,Pillepich:2006fj,Munoz:2015eqa,Shimabukuro:2015iqa}, lensing \cite{Kovetz:2012jq}, the late ISW effect \cite{Raccanelli:2015lca}, and interactions between baryons and dark matter \cite{Tashiro:2014tsa,Munoz:2015bca}.
Here, we are interested in measuring rotational asymmetries; being able to access small angular scales gives the opportunity to distinguish the dipolar asymmetry generated by a variable spectral index, surpassing the intermediate scales at which it would vanish.
We here compute the angular power spectrum of 21-cm fluctuations sourced by the dipolar and quadrupolar asymmetries, including several nontrivial scale dependencies motivated by theories and observations as mentioned above.
By the simple application of an estimator for CMB rotational asymmetry \cite{Pullen:2007tu,Hanson:2009gu}, we forecast how well different 21-cm surveys can constrain departures from rotational invariance.

We structure this paper as follows. In Sec.~\ref{sec:21cm} we introduce the 21-cm line observable and its power spectrum. Then, in Sec.~\ref{sec:ani}, we present theoretical models for breaking rotational invariance and show their predictions. In Sec.~\ref{sec:results} we show our results, and we conclude in Sec.~\ref{sec:conclusions}.

\section{21-cm cosmology}
\label{sec:21cm}

The nuclear spin of the hydrogen atom makes its triplet ground state have a slightly higher energy than its singlet counterpart, giving rise to a transition with characteristic wavelength $\lambda \approx 21 \ {\rm cm}$ in the radio spectrum. Its very long wavelength makes it a good probe of the early Universe, being easily identifiable.
We start by reviewing the physics of the 21-cm line during the dark ages and its angular power spectrum.

\subsection{Global signal}

The ratio of the populations of the triplet and singlet hydrogen states defines a temperature, which we denote as spin temperature $T_s$. During the dark ages, CMB photons stimulate radiative transitions between the singlet and the triplet hydrogen states \cite{Loeb:2003ya}. Collisions between different hydrogen atoms will also create upwards and downwards transitions. The time scale of both these effects is much smaller than the evolution of the Universe \cite{Lewis:2007kz}, so we can use the quasisteady-state approximation,
\be
n_0(C_{01}+R_{01}) = n_1(C_{10}+R_{10}),
\ee
where $n_1$ and $n_0$ are the densities of triplet- and singlet-state hydrogen atoms, $C_{ij}$ are the collisional transition rates, and $R_{ij}$ are the rates of radiative transition due to the CMB blackbody photons. This allows us to define the spin temperature, which we can approximate very well by
\be
T_s = T_\gamma + \dfrac{C_{10}}{C_{10}+A_{10} \frac{T_H}{T_*}}(T_H - T_\gamma),
\ee
where $T_H$ is the temperature of the neutral hydrogen, $T_{\gamma}$ is that of the CMB, $T_*=0.068 \, {\rm K} = 5.9 \, \mu{\rm eV}$ is the characteristic temperature of the 21-cm transition, and $A_{10}$ is the Einstein spontaneous emission coefficient of the 21-cm transition.

During the redshift period of interest, collisions dominate over radiative transitions, which couples the spin temperature to that of the hydrogen. 
This enables hydrogen atoms to resonantly absorb CMB photons with a rest wavelength of 21 cm, which from Earth results in a decrease in the brightness temperature of the CMB at the corresponding redshifted wavelength. We ignore here low-redshift effects, such as the Wouthuysen-Field effect \cite{Field,Wouthuysen,Hirata:2005mz}, heating of the hydrogen gas due to miniquasars \cite{Kuhlen:2005es,Furlanetto:2006jb}, or early stellar formation \cite{Fialkov:2012su,Haiman:2003he}.

Let us define the 21-cm line temperature for small optical depths $\tau$ as
\be
T_{21} = \dfrac{T_s-T_\gamma}{1+z}\tau,
\ee
corresponding to the contrast with the CMB temperature redshifted to today. The optical depth is given by
\be
\tau = \dfrac{3}{32 \pi} \dfrac{T_*}{T_s}n_{H} \lambda_*^3 \dfrac{A_{10}}{H(z)+(1+z)\partial_rv_r},
\ee
where $\lambda_* \approx 21$ cm, $n_{H}$ is the number density of neutral hydrogen, and $\partial_rv_r$ is the proper gradient of the peculiar velocity along the line of sight.

\subsection{Fluctuations}

The optical depth and the spin temperature of a hydrogen clump depend on its density and velocity divergence. The small anisotropies in these two quantities create fluctuations on the 21-cm temperature $T_{21}$. Let us define $\delta_v\equiv -(1+z) (\partial_r v_r) / H(z)$. Then, at linear order, the 21-cm fluctuations can be expressed as \cite{Ali-Haimoud:2013hpa,Munoz:2015eqa}
\begin{eqnarray}
\delta T_{21}({\bf x})
&=&  \alpha(z) \delta_b({\bf x}) + \bar T_{21}(z) \delta_v({\bf x}),
\label{eq:T21_3D}
\end{eqnarray}
with $\bar T_{21}(z)$ being the spatially averaged 21-cm brightness temperature, and $\alpha(z)= {\rm d} T_b / {\rm d} \delta_b$, including gas temperature fluctuations as in Ref.~\cite{Munoz:2015eqa}.
The observed fluctuation of this quantity in a direction $\hat n$ of the sky and at a certain frequency $\nu$ is given by
\begin{eqnarray}
\delta T_{21}(\hat{n}, \nu) = \int_0^{\infty} dx W_\nu(x) \delta T_{21}({\bf x}) ~, \label{eq:T21_2D}
\end{eqnarray}
where $W_\nu(x)$ is the window function selecting the information at a certain frequency band centered in $\nu$. In Fourier space, the primordial curvature perturbation $\zeta_{\bf k}$ is related to the baryonic anisotropies by $\delta_b({\bf k}, z) =  M_\zeta(k,z)\zeta_{\bf k}$ and $\delta_v({\bf k}, z)=\mu^2\delta_b({\bf k}, z)$, with $\mu=(\hat k \cdot \hat n)$. We can, therefore, define the transfer function of the 21-cm temperature fluctuations as
\begin{eqnarray}
{\cal T}_{\ell}(k,\nu) 
= \int_0^{\infty} dx W_\nu(x) M_\zeta[k,z(x)] \left [\bar{T}_{21}(z) {\cal J}_\ell(kx) + \alpha(z) j_\ell(kx) \right],
\end{eqnarray}
where $j_\ell$ is the spherical Bessel function with index $\ell$, and we have defined ${\cal J}_\ell (kx) \equiv - \partial^2 j_\ell(kx) / (\partial kx)^2$, which can be written in terms of $j_\ell$, and $j_{\ell\pm2}$ \cite{Bharadwaj:2004nr}. Given this, we can easily compute the 21-cm line angular power spectrum at a certain frequency $\nu$ as
\be
C_{\ell} = \frac{2}{\pi}
\int_0^\infty k^2 dk P_0(k) 
{\cal T}_{\ell}^2(k,\nu) \; 
\label{eq:Cell} ,
\ee
where $P_0(k)$ is the (isotropic) primordial curvature power spectrum, given in $\Braket{\zeta_{{\bf k}}^{\rm iso} \zeta_{{\bf k}'}^{\rm iso}} = (2\pi)^3 P_0(k) \delta^{(3)}\left({\bf k} + {\bf k}' \right)$, with $\zeta_{\bf k}^{\rm iso}$ being the isotropic part of the curvature perturbations.

Finally, we can define the mode-coupling matrix as
\begin{eqnarray}
G_{\ell_1 \ell_2}
= \frac{2}{\pi}
\int_0^\infty k^2 dk P_0(k) f(k)
{\cal T}_{\ell_1}(k,\nu)  
{\cal T}_{\ell_2}(k,\nu) ,
\label{eq:Gfunc} 
\end{eqnarray}
where the additional factor $f(k)$ is 1 in the usual case and will have different values for alternative inflationary models. 
This is the standard general way to express the angular correlations; in the following section, we compute the angular power spectra for different models describing rotational asymmetries.

For computational efficiency, we employ the flat-sky approximation \cite{Hu:2000ee,White:1999uk} for $\ell\geq 10^3$ for a bandwidth of $\Delta \nu = 1.0 \, {\rm MHz}$ and $\ell\geq 10^4$ for a bandwidth of $0.1 \, {\rm MHz}$. In that approximation, it is easy to check that $G_{\ell_1 \ell_1+\Delta\ell}=G_{\ell_1 \ell_1}$, up to $O(\fsky)$ factors that we ignore (see the Appendix \ref{appen:flat-sky}). 

\subsection{Instrumental Noise}

In the cosmic variance limit (CVL), the only source of noise is the variance given by having a finite number of measurements of the power spectrum $C_\ell$ itself. If one considers, however, an interferometer looking at the dark ages at a certain frequency $\nu$, there is an additional noise power spectrum given by \cite{Jaffe:2000yt,Knox:2002pe,Kesden:2002ku,Zaldarriaga:2003du}
\be
\ell^2 C_\ell^N = \dfrac{(2\pi)^3 T_{\rm sys}^2(\nu)}{\Delta\nu \,t_o f_{\rm cover}^2}\left( \dfrac{\ell}{\ell_{\rm cover}(\nu)}\right)^2,
\ee
where $\Delta\nu$ is the bandwidth of the survey, $t_o$ is the total time of observation, $\ell_{\rm cover}(\nu)\equiv 2\pi D_{\rm base}/\lambda$ is the maximum multipole observable, with $D_{\rm base}$ being the largest baseline of the interferometer. The amplitude of this noise is given by the system temperature $T_{\rm sys}$, which we take to be the synchrotron temperature of the observed sky
\be
T_{\rm sys}(\nu) = 180 \left( \dfrac{\nu}{180 \, \rm MHz}\right)^{-2.6} \rm K,
\ee
found from extrapolating to lower frequencies the results in Ref.~\cite{Bowman:2008mk}.

In this paper, we consider three different noise levels. First, we use specifications consistent with those of the Square Kilometre Array~\cite{ska}, currently under construction in South Africa and Australia (note that there has recently been a ``rebaselining'' of the planned instrument, and exact specifications and survey strategies are still somewhat in development; we perform our calculations by assuming instrument specifications specified below) . We then assume a futuristic radio array (FRA), as an example for a futuristic, but still Earth-based, experiment. Finally, we show results for the cosmic variance limited (CVL) case, where $C_\ell^N = 0$, to show, as a proof of principle, the theoretical limits that can be obtained by using this probe.

For the SKA case, we take a baseline of 6 km, a coverage fraction of $f_{\rm cover}=0.02$, and a time of data collection of 5 whole years. In the FRA case, we consider an increased baseline of $D_{\rm base} = 100 \ {\rm km}$, a coverage fraction of $0.2$, and 10 whole years of observations. The baseline of this FRA case would be large enough to reach $\ell\gtrsim 10^4$ at redshift $z \lesssim 50$, enabling us to prove the small-scale anisotropies predicted by some models \cite{Dai:2013kfa}.


\section{Rotational asymmetries}
\label{sec:ani}

  The CMB measurements indicate that the microwave sky is almost perfectly rotationally symmetric, while there are a few observed anomalies that require explaining, the most important being the dipolar asymmetry. Data analyses of the CMB fluctuations \cite{Hansen:2004vq, Gordon:2005ai, Eriksen:2007pc, Gordon:2006ag, Ade:2013nlj,Akrami:2014eta,Ade:2015lrj,Ade:2015sjc,Aiola:2015rqa} and the density of quasars at lower redshifts \cite{Hirata:2009ar} indicate the existence of the dipolar asymmetry at very large scales and its decaying behavior up to $k \sim 1 \ {\rm Mpc^{-1}}$. There are also upper bounds on the quadrupolar asymmetry \cite{Kim:2013gka,Ade:2015lrj,Ade:2015sjc}, which can be generated by a wide range of inflation models involving anisotropic sources, such as vector fields. The signatures of these anomalies may be discovered in observables other than CMB. 

  In this section, we investigate the angular power spectrum of the 21-cm fluctuations, $\Braket{a_{\ell_1 m_1} a_{\ell_2 m_2}}$ generated by such anomalies. We then show that nonvanishing off-diagonal components satisfying $|\ell_1 - \ell_2| = 1$ and $2$ appear in the dipolar and quadrupolar asymmetry case, respectively, as unique signatures of statistical homogeneity and isotropy violation. We begin with the dipolar asymmetry.

\subsection{Primordial dipolar asymmetry}\label{sec:dipole}

Let us consider a curvature perturbation with a position-dependent dipolar modulation written as
\begin{eqnarray}
\zeta_{\bf k}({\bf x}) = \zeta_{\bf k}^{\rm iso}
\left[1 + \sum_M A_{1M} f(k) Y_{1M}(\hat{n}) \right] ~, \label{eq:zeta_A1M}
\end{eqnarray}
where ${\bf x} \equiv x \hat{n}$ with $x \equiv |{\bf x}|$. A constant $f(k) = 1$ was originally introduced to explain a $\sim 3\sigma$  evidence of dipolar asymmetry in the CMB sky at very large scales ($\ell \lesssim 60$) \cite{Ade:2013nlj,Akrami:2014eta,Ade:2015lrj,Ade:2015sjc,Aiola:2015rqa}. This asymmetry can be expressed as an amplitude $A \sim 0.07$ in $T(\hat{n}) = T^{\rm iso}(\hat{n})(1 + A \hat{n} \cdot \hat{p})$, with $\hat{p}$ denoting {the} preferred direction of the modulation, and $T^{\rm iso}$ being the usual temperature fluctuation. On the other hand, several detailed analyses of the CMB maps have also shown that such dipolar modulation is highly damped for $\ell \gtrsim 60$ \cite{Ade:2013nlj,Akrami:2014eta,Ade:2015lrj,Ade:2015sjc,Aiola:2015rqa}. This scaling seems to be consistent with a different constraint obtained by quasar abundances, leading to a vanishing dipolar asymmetry at $k \sim 1 \ {\rm Mpc}^{-1}$ \cite{Hirata:2009ar}. We implement a heuristic model for such observationally motivated scale dependence as a function, {$f(k) = (1-k/k_A)^n$ with $k_A \equiv 1 \ {\rm Mpc}^{-1}$}, for $n=1$ and 2, in our parametrization of Eq.~\eqref{eq:zeta_A1M}. There are also theoretical motivations for this $f(k)$ shape, such a spatially varying spectral index $n_s$ \cite{Dai:2013kfa}, where the asymmetry would cancel at some scale $k_A$.​

Let us compute the spherical harmonic coefficient $a_{\ell m}$ of the 21-cm fluctuations from $\delta T_{21}(\hat{n},\nu) = \sum_{\ell m} a_{\ell m}(\nu) Y_{\ell m}(\hat{n})$. We can separate $a_{\ell m}$, due to Eq.~\eqref{eq:zeta_A1M}, in an isotropic and dipolar part as
\begin{eqnarray}
a_{\ell m}(\nu) = a_{\ell m}^{\rm iso}(\nu) + a_{\ell m}^{\rm dip}(\nu) ~,
  \end{eqnarray}
where the isotropic part is formulated in a usual way, reading
\begin{eqnarray}
  a_{\ell m}^{\rm iso}(\nu) =  4\pi i^\ell 
  \int \frac{d^3 k}{(2\pi)^3} Y_{\ell m}^*(\hat{k}) \zeta_{\bf k}^{\rm iso} 
 {\cal T}_{\ell}(k,\nu) ~, \label{eq:alm21_iso}
  \end{eqnarray}
 and the part due to the dipolar asymmetry is
\begin{eqnarray}
a_{\ell m}^{\rm dip}(\nu)
&=&
\sum_{LM} 4\pi i^L 
\int \frac{d^3 k}{(2\pi)^3}
Y_{LM}^*(\hat{k}) \zeta_{\bf k}^{\rm iso} f(k)
{\cal T}_L^{}(k, \nu)
\times
(-1)^m
h_{\ell L1}
\sum_{M'} A_{1M'} 
\left(
\begin{array}{ccc}
\ell & L & 1 \\
-m & M & M'
\end{array}
\right) ~,
\end{eqnarray}
where 
\begin{eqnarray}
h_{l_1 l_2 l_3} 
&\equiv& \sqrt{\frac{(2 l_1 + 1)(2 l_2 + 1)(2 l_3 + 1)}{4 \pi}}
\left(
  \begin{array}{ccc}
  l_1 & l_2 & l_3 \\
   0 & 0 & 0
  \end{array}
 \right)~. 
\end{eqnarray}
Assuming a small dipolar asymmetry amplitude, i.e., $|A_{1M} f(k)| \ll 1$, we can treat it perturbatively, {thus obtaining} the angular correlations of $a_{\ell m}$ as
\begin{eqnarray}
  \Braket{a_{\ell_1 m_1} a_{\ell_2 m_2}}
  &\simeq& \Braket{a_{\ell_1 m_1}^{\rm iso} a_{\ell_2 m_2}^{\rm iso}}
  + \Braket{a_{\ell_1 m_1}^ { \rm iso} a_{\ell_2 m_2}^{ \rm dip}}
  + \Braket{a_{\ell_1 m_1}^{ \rm dip} a_{\ell_2 m_2}^{ \rm iso}}
  \nonumber \\
  &=& C_{\ell_1} (-1)^{m_1} \delta_{\ell_1, \ell_2} \delta_{m_1, -m_2}
  + (-1)^{m_2} C_{\ell_1 m_1, \ell_2 -m_2} ~,
\end{eqnarray}
where $C_\ell$ is the same as Eq.~\eqref{eq:Cell} and we have defined
\begin{eqnarray}
C_{\ell_1 m_1, \ell_2 m_2} 
= 
\left( G_{\ell_1 \ell_1} + G_{\ell_2 \ell_2}  \right) 
  (-1)^{m_1} h_{\ell_1 \ell_2 1}
  \sum_{M} A_{1M} 
  \left(
  \begin{array}{ccc}
    \ell_1 & \ell_2 & 1 \\
    -m_1 & m_2 & M
  \end{array}
  \right) ~, \label{eq:Cl1l2_A1M}
\end{eqnarray}
using the mode-coupling matrix $G_{\ell_1 \ell_2}$, defined in Eq.~\eqref{eq:Gfunc}. Note that due to parity conservation and the triangular inequalities of $h_{\ell_1 \ell_2 1}$, the only nonzero signals of $C_{\ell_1 m_1, \ell_2 m_2}$ occur at $\ell_1 = \ell_2 \pm 1 $. This has the identical function form to the correlation of CMB fluctuations \cite{Pullen:2007tu}.

While the precise scale dependence is different for different models \cite{Schmidt:2012ky, Dai:2013kfa,Lyth:2013vha,Abolhasani:2013vaa,Jazayeri:2014nya}, we choose the following two shapes as proxies:
\begin{eqnarray}
  f(k) = 1 - \frac{k}{k_A} ~, \ \
  \left(1 - \frac{k}{k_A}\right)^2 ~. \label{eq:fk_A1M}
  \end{eqnarray}
These reconstruct the observed decaying shapes for $k < k_A (\equiv 1 \ {\rm Mpc^{-1}})$.\footnote{Strictly speaking, the CMB constraints obtained from $100 \lesssim \ell \lesssim 1000$ \cite{Ade:2013nlj,Akrami:2014eta,Ade:2015lrj,Ade:2015sjc,Aiola:2015rqa} favor a bit more rapidly decaying behaviors than Eq.~\eqref{eq:fk_A1M}, for $0.01 \ {\rm Mpc^{-1}} \lesssim k \lesssim 0.1 \ {\rm Mpc^{-1}} $.} On the other hand, on unobserved small scales as $k > k_A$, these grow larger and with opposite signs.

\begin{figure}[t]
\begin{tabular}{cc}
    \begin{minipage}[t]{0.5\hsize}
  \begin{center}
    \includegraphics[width=1\textwidth]{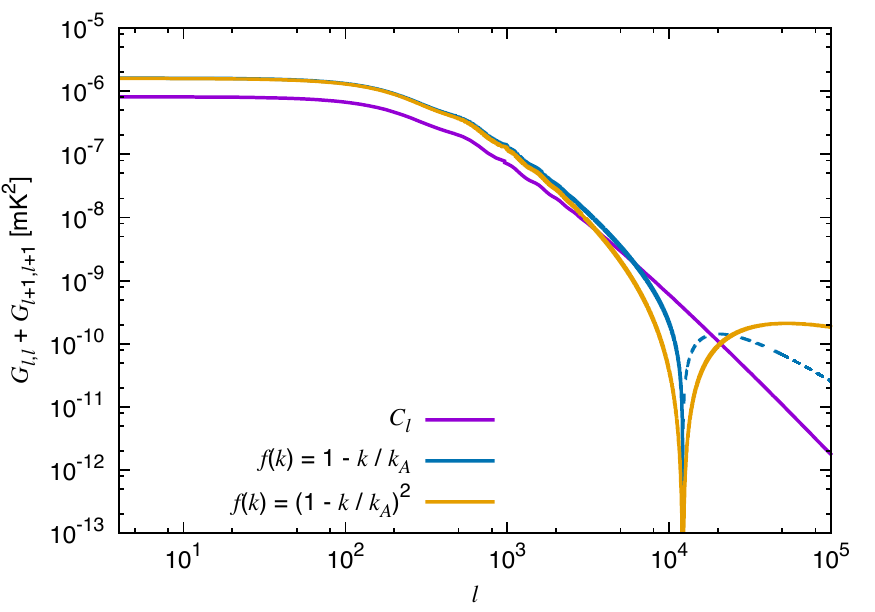}
  \end{center}
    \end{minipage}
\begin{minipage}[t]{0.5\hsize}
  \begin{center}
    \includegraphics[width=1\textwidth]{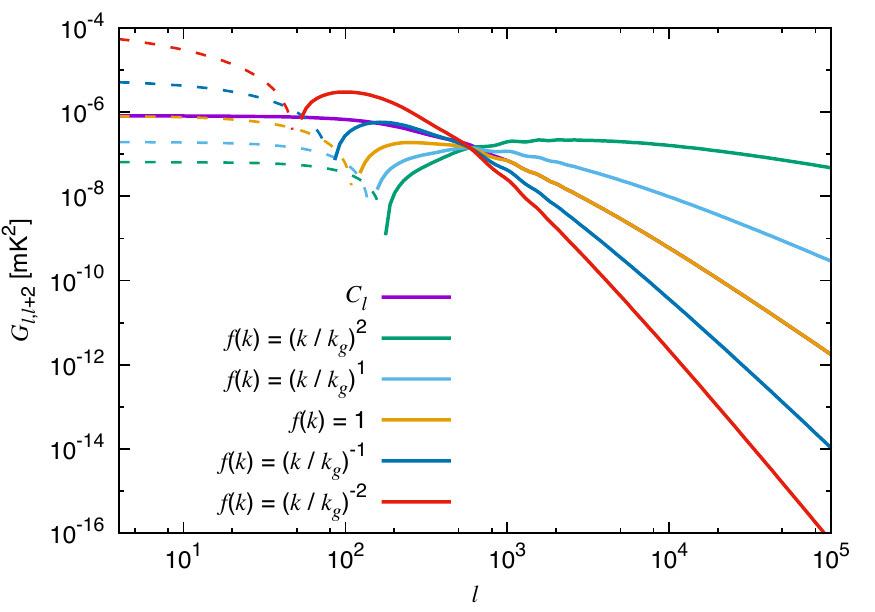}
  \end{center}
\end{minipage}
\end{tabular}
\caption{Correlations $G_{\ell, \ell} + G_{\ell + 1, \ell + 1}$ for $A_{1M}$ (left panel) and $G_{\ell, \ell + 2}$ for $g_{2M}$ (right panel). Solid (dashed) lines describe positive (negative) values. We here adopt $z = 30$ and $\Delta \nu = 1.0 \, {\rm MHz}$. For the pivot scales, we choose $k_A = 1 \ {\rm Mpc^{-1}}$ in the left panel and $k_g = 0.05 \ {\rm Mpc^{-1}}$ in the right panel, corresponding to $\ell_A \equiv k_A x(z = 30) \simeq 1.3 \times 10^4$ and $\ell_g \equiv k_g x(z = 30) \simeq 650$ in $\ell$ space, respectively. For comparison, we also show the isotropic power spectrum $C_\ell = G_{\ell \ell}^{f(k) = 1}$ (purple), which acts as a cosmic variance in the Fisher matrix computation.}
\label{fig:Cls}
\end{figure}

Figure~\ref{fig:Cls} plots the correlations $G_{\ell, \ell} + G_{\ell + 1, \ell + 1}$ for the two different models for the function $f(k)$, taking $z = 30$ and $\Delta \nu = 1.0 \, {\rm MHz}$. We can confirm there that a dip is located at $\ell \sim \ell_A \equiv  k_A x(z = 30) \simeq 1.3 \times 10^4$, with $x(z = 30) \simeq 13 \, {\rm Gpc}$ denoting the conformal distance to $z=30$, as expected. The difference between these two models show up for $\ell \gtrsim \ell_A$ because of the difference of the signs and the different power law exponent. We also notice that {$G_{\ell, \ell} + G_{\ell + 1, \ell + 1}$ is substantially enhanced compared with the isotropic power spectrum $C_\ell$}, which we also show in Fig.~\ref{fig:Cls} for reference, for $\ell \gtrsim \ell_A$, as expected from Eq.~\eqref{eq:fk_A1M}. This results in a sharp rise in the signal-to-noise ratio for $\ell \gtrsim \ell_A$, as we show. Notice also that the angular behavior of $G_{\ell_1 \ell_2}$ is heavily dependent on $f(k)$, which could help to disentangle the information about asymmetries in the data analysis of the 21-cm power spectrum.

\subsection{Primordial quadrupolar asymmetry}\label{sec:quad}

When including a quadrupolar asymmetry, we can write the curvature power spectrum as
\begin{eqnarray}
  \Braket{\zeta_{{\bf k}_1} \zeta_{{\bf k}_2}}
  &=& (2\pi)^3 P_0(k_1) 
  \left[ 1 + \sum_{M} g_{2M} f(k_1) Y_{2M}(\hat{k}_1) \right]
  \delta^{(3)}({\bf k}_1 + {\bf k}_2)~. \label{eq:zeta2_g2M}
\end{eqnarray}
A nonvanishing $g_{2M}$ arises in inflation models where the inflaton couples to a vector field with a nonzero vacuum expectation value via {${\cal L} = - \frac{1}{4}I^2(\phi) F^2$} \cite{Dimastrogiovanni:2010sm,Soda:2012zm,Maleknejad:2012fw,Bartolo:2012sd,Naruko:2014bxa,Bartolo:2014hwa,Bartolo:2015dga,Abolhasani:2015cve}. In these cases, the time dependence of the coupling function $I(\phi)$ determines the scale dependence of $f(k)$. The nearly scale-invariant spectrum, i.e., $f(k) = 1$, is realized by choosing $I(\phi) \propto a^{-2}$, with $a$ denoting the scale factor \cite{Bartolo:2012sd,Naruko:2014bxa,Bartolo:2015dga}. A nearly scale-invariant $f(k)$ is also generated in the solid inflation models \cite{Bartolo:2013msa, Bartolo:2014xfa}.
Most other models predict a scale dependence for $f(k)$.  
The magnitude of $g_{2M}$ relies strongly on the model parameters. For vector field models, $g_{2M}$ is proportional to the ratio of the energy density of the vector field to that of the scalar field, $\rho_A / \rho_\phi$, \cite{Bartolo:2012sd,Naruko:2014bxa}. If the inflaton is identified to the pseudoscalar field, the coupling strength between the pseudoscalar and the vector field also affects $g_{2M}$ \cite{Bartolo:2014hwa,Bartolo:2015dga,Abolhasani:2015cve}. The data analysis with the {\it Planck} map gives upper bounds for the scale invariant case $f(k) = 1$ of $|g_{2M}| \lesssim 10^{-2}$ \cite{Kim:2013gka, Ade:2015lrj,Ade:2015sjc}, leading to several constraints on the model parameters, e.g. $\rho_A / \rho_\phi \lesssim 10^{-9}$ \cite{Bartolo:2015dga}.

The angular correlations generated from Eq.~\eqref{eq:zeta2_g2M} then become
\begin{eqnarray}
  \Braket{a_{\ell_1 m_1} a_{\ell_2 m_2}}
  = C_{\ell_1} (-1)^{m_1} \delta_{\ell_1, \ell_2} \delta_{m_1, -m_2}
  + (-1)^{m_2} C_{\ell_1 m_1, \ell_2 -m_2}
\end{eqnarray}
where $C_\ell$ is given by Eq.~\eqref{eq:Cell} and
\begin{eqnarray}
  C_{\ell_1 m_1, \ell_2 m_2}
  &=& i^{\ell_1 - \ell_2}
  G_{\ell_1 \ell_2}
 (-1)^{m_1} 
h_{\ell_1 \ell_2 2}
\sum_{M}  g_{2M} 
\left(
  \begin{array}{ccc}
  \ell_1 & \ell_2 & 2 \\
   -m_1 & m_2 & M 
  \end{array}
 \right) ~. \label{eq:Cl1l2_g2M}
\end{eqnarray}
Likewise to the dipolar case, due to  the triangular inequalities of $h_{\ell_1 \ell_2 2}$ and parity, $C_{\ell_1 m_1, \ell_2 m_2}$ vanishes except for $|\ell_1 - \ell_2| = 0, 2$. An identical functional form is seen in CMB off-diagonal correlations \cite{Pullen:2007tu}.

  In what follows, we discuss the detectability of $g_{2M}$ in a general model-independent way and do not translate the results into any specific model parameters. To do so, we assume five different power law shapes for $f(k)$,
  \begin{eqnarray}
    f(k) = 1 ~, \ \
    \left( \frac{k}{k_g} \right)^{\pm 1}  ~, \ \
    \left( \frac{k}{k_g} \right)^{\pm 2}  ~,  \label{eq:fk_g2M}
  \end{eqnarray}
  with $k_g = 0.05 \ {\rm Mpc^{-1}}$ being the pivot scale adopted in the {\it Planck} collaboration \cite{Ade:2015lrj,Ade:2015sjc}. 

  The right panel of Fig.~\ref{fig:Cls} plots the off-diagonal correlations $G_{\ell, \ell + 2}$ generated from the five different models in Eq.~\eqref{eq:fk_g2M} for $z = 30$ and $\Delta \nu = 1.0 \, {\rm MHz}$. It is obvious in this figure that all the lines intersect each other at the multipole corresponding to the pivot scale, $\ell_g \equiv k_g x(z = 30) \simeq 650$, and they are tilted depending on their spectral indices. We can also observe sign changes at $\ell \sim 100$.


\section{Results}\label{sec:results}

Here, we forecast measurements of rotational asymmetries with the 21-cm power spectrum. We compute $G_{\ell_1 \ell_2}$ for the seven different models of the function $f(k)$ (two for the dipolar and five for the quadrupolar asymmetry), and then calculate the forecasted detectability of the coefficients $A_{1M}$ and $g_{2M}$ via a Fisher matrix analysis, including the three different instrumental noise spectra of Sec.~\ref{sec:21cm}: $C_\ell^N = 0$ (CVL), $C_\ell^N$ in SKA and $C_\ell^N$ in a futuristic radio array.

\begin{figure}[t]
  \begin{tabular}{cc}
    \begin{minipage}{0.5\hsize}
  \begin{center}
    \includegraphics[width=1\textwidth]{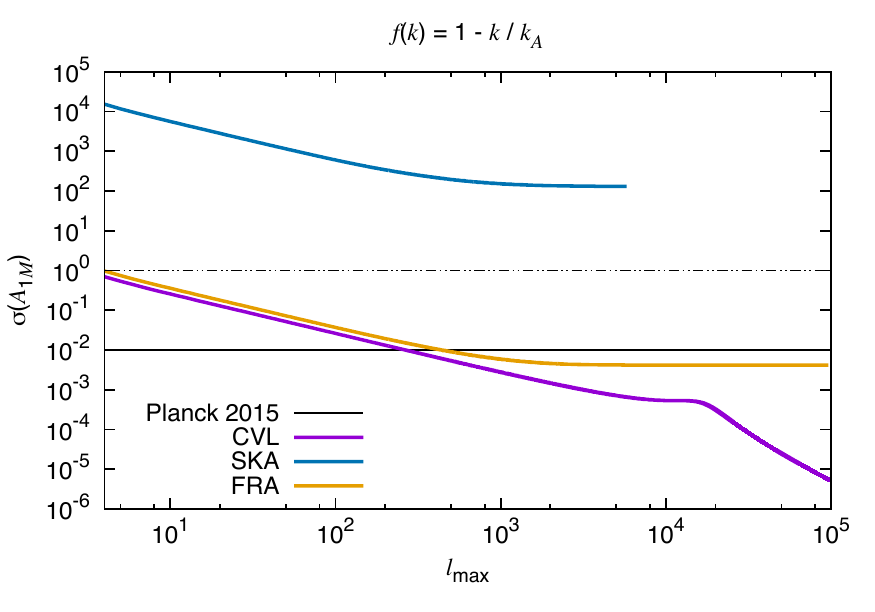}
  \end{center}
\end{minipage}
\begin{minipage}{0.5\hsize}
  \begin{center}
    \includegraphics[width=1\textwidth]{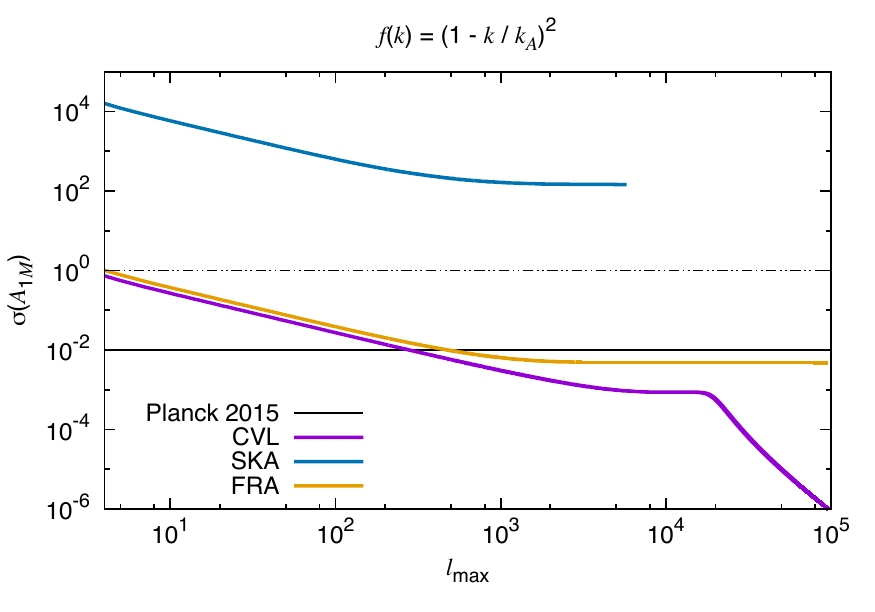}
  \end{center}
\end{minipage}
\end{tabular}
  \\
    \begin{tabular}{c}
    \begin{minipage}{1.0\hsize}
  \begin{center}
    \includegraphics[width = 0.5\textwidth]{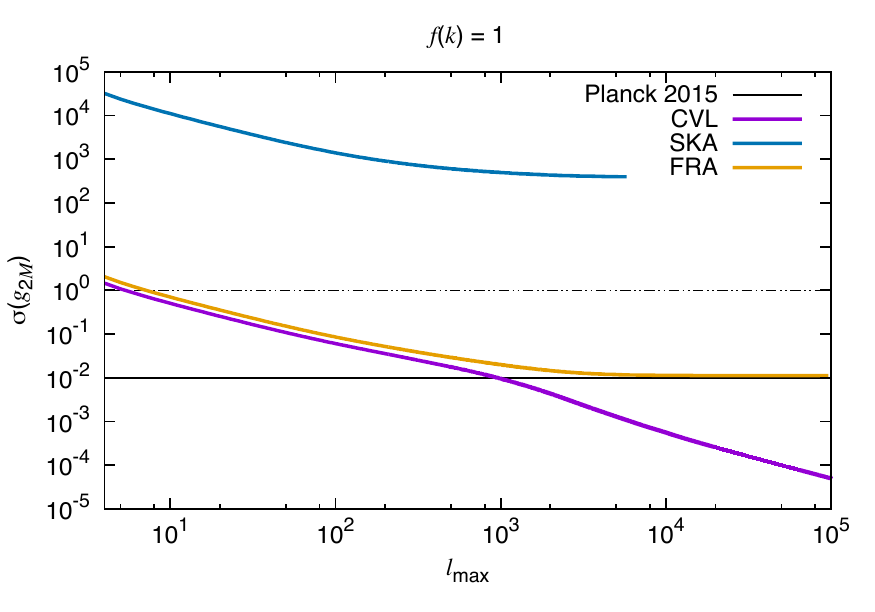}
  \end{center}
\end{minipage}
\end{tabular}
\\
  \begin{tabular}{cc}
    \begin{minipage}{0.5\hsize}
  \begin{center}
    \includegraphics[width=1\textwidth]{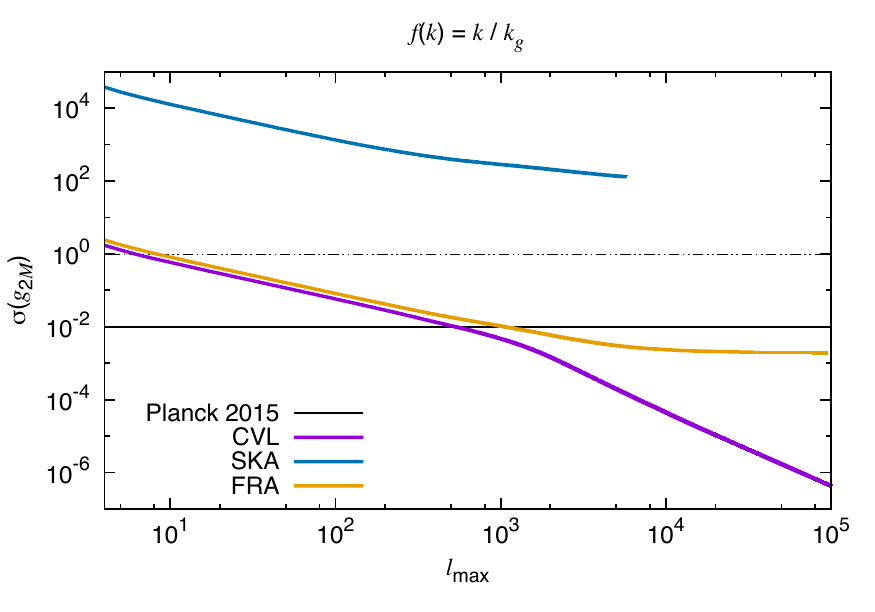}
  \end{center}
\end{minipage}
\begin{minipage}{0.5\hsize}
  \begin{center}
    \includegraphics[width=1\textwidth]{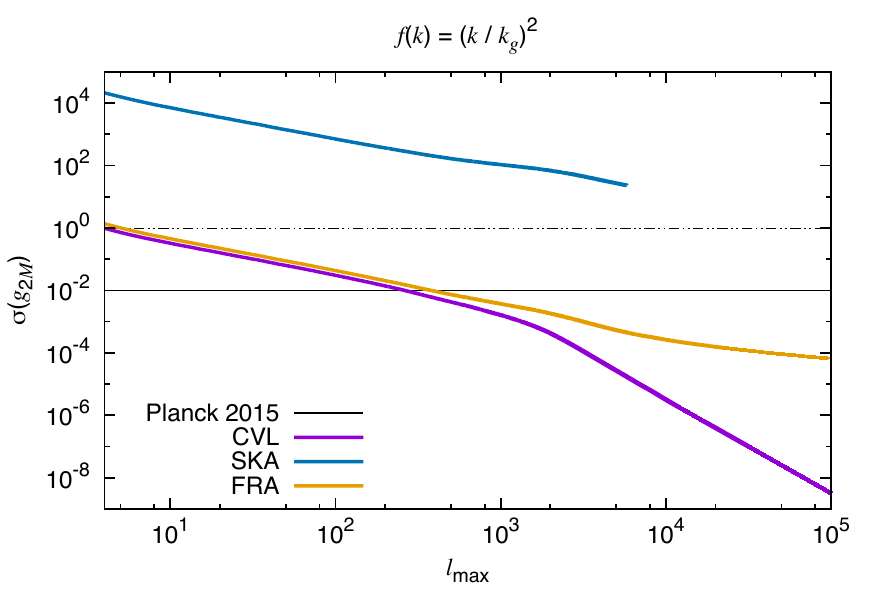}
  \end{center}
\end{minipage}
\end{tabular}
  \caption{Expected $1\sigma$ errors on $A_{1M}$ and $g_{2M}$ for a single redshift slice at $z = 30$ and $\Delta \nu = 0.1 \, {\rm MHz}$ with $\ell_{\rm min} = 2$. We show the results for $f(k)=(1-k/k_A)^{1,2}$ for the dipolar case and for $f(k)=(k/k_g)^{0,1,2}$ for the quadrupolar case. For comparison, we draw the $1\sigma$ errors obtained from the {\it Planck} 2015 bounds, i.e., $\sigma(A_{1M}) \sim \sigma(g_{2M}) \sim 0.01$ \cite{Kim:2013gka,Ade:2015sjc,Ade:2015lrj}. We here truncate larger $\ell$ modes than the baseline sizes: $\ell_{\rm cover} = 5791$ for SKA and $96516$ for FRA.}
\label{fig:err}
\end{figure}

\begin{figure}[t]
  \begin{tabular}{cc}
    \begin{minipage}{0.5\hsize}
  \begin{center}
    \includegraphics[width=1\textwidth]{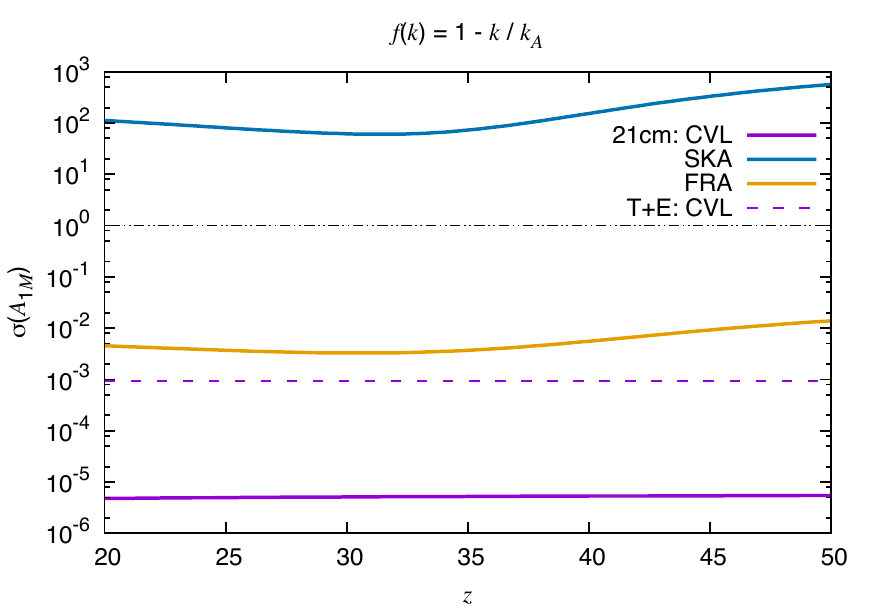}
  \end{center}
\end{minipage}
\begin{minipage}{0.5\hsize}
  \begin{center}
    \includegraphics[width=1\textwidth]{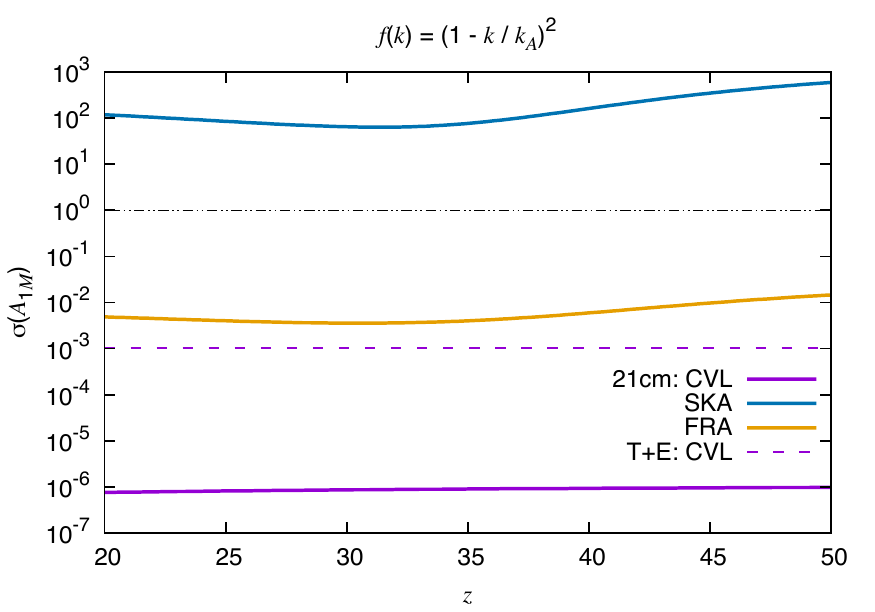}
  \end{center}
\end{minipage}
\end{tabular}
  \\
    \begin{tabular}{c}
    \begin{minipage}{1.0\hsize}
  \begin{center}
    \includegraphics[width = 0.5\textwidth]{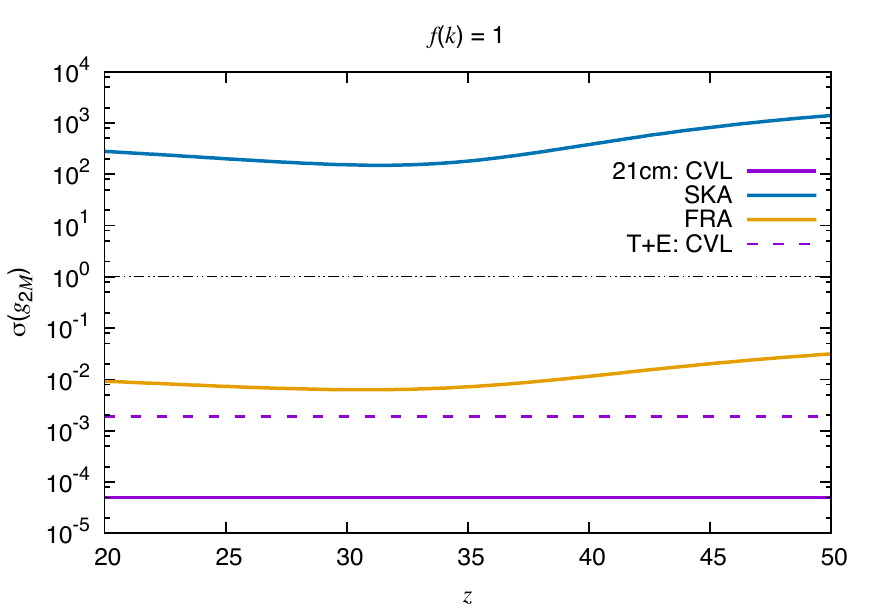}
  \end{center}
\end{minipage}
\end{tabular}
\\
  \begin{tabular}{cc}
    \begin{minipage}{0.5\hsize}
  \begin{center}
    \includegraphics[width=1\textwidth]{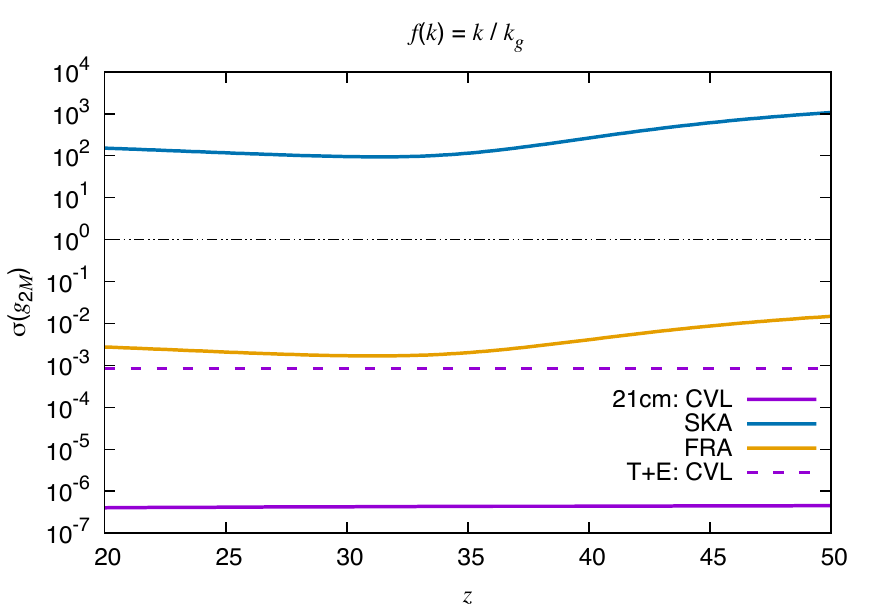}
  \end{center}
\end{minipage}
\begin{minipage}{0.5\hsize}
  \begin{center}
    \includegraphics[width=1\textwidth]{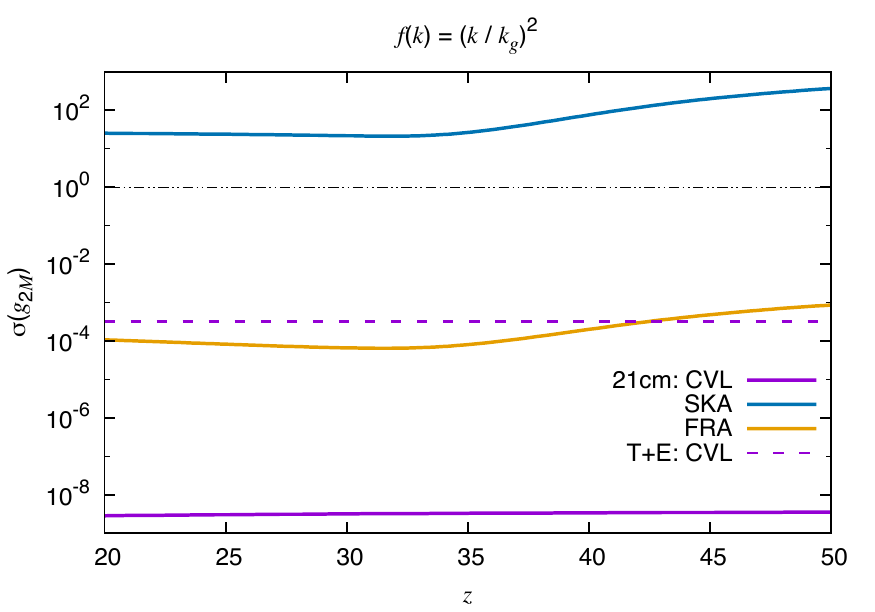}
  \end{center}
\end{minipage}
\end{tabular}
\caption{Expected $1\sigma$ errors on $A_{1M}$ and $g_{2M}$ for a single redshift slice adding up all modes up to $\ell_{\rm max} = {\rm Min}[10^5, \ell_{\rm cover}]$ as a function of redshift $z$. We here take $\Delta \nu = 1.0 \, {\rm MHz}$ and $\ell_{\rm min} = 2$. For comparison, we also show the expected errors we obtain by analyzing CMB temperature and E-mode polarization jointly (T+E) in a noiseless (CVL) CMB measurement up to $\ell = 2000$.}
\label{fig:err_dip-quad_z}
\end{figure}

  Since the functional forms of $C_{\ell_1 m_1 \ell_2 m_2}$ for 21-cm and CMB fluctuations are the same, we employ an optimal CMB estimator for the asymmetric parameters $h_{LM} \in (A_{1M}, g_{2M})$ \cite{Pullen:2007tu,Hanson:2009gu, Hanson:2010gu,Ma:2011ii} in our 21-cm analysis, expressed as
\begin{eqnarray}
\hat{h}_{LM} &=& \frac{1}{2} \sum_{L'M'} F_{LM,L'M'}^{-1}
\sum_{\ell_1 m_1 \ell_2 m_2} \bar{a}^{ *}_{\ell_1 m_1}
\frac{\partial C_{\ell_1 m_1 \ell_2 m_2}}{\partial h_{L'M'}^{*}} 
 \bar{a}^{}_{\ell_2 m_2} ~,
\end{eqnarray} 
where $\bar{a}^{}_{\ell m}$ are the harmonic coefficients weighted with the inverse of the covariance matrix, and the Fisher matrix is given as  
\begin{eqnarray}
F_{LM, L' M'}
  = \frac{1}{4} 
  \sum_{\substack{\ell_1 m_1 \ell_2 m_2 \\ \ell_1' m_1' \ell_2' m_2'}}
\frac{\partial C_{\ell_1 m_1 \ell_2 m_2}}{\partial h_{L M}^{*}} 
\frac{\partial C_{\ell_1' m_1' \ell_2' m_2'}^{*}}{\partial h_{L'M'}} 
\left( \Braket{\bar{a}^{ *}_{\ell_1 m_1} \bar{a}^{}_{\ell_2 m_2}
  \bar{a}^{}_{\ell_1' m_1'} \bar{a}^{ *}_{\ell_2' m_2'} }
- \Braket{\bar{a}^{ *}_{\ell_1 m_1} \bar{a}^{}_{\ell_2 m_2}}
 \Braket{ \bar{a}^{}_{\ell_1' m_1'} \bar{a}^{ *}_{\ell_2' m_2'} }
 \right) ~.
 \end{eqnarray}
For simplicity, let us work with the diagonal approximation of the inverse of the covariance matrix. Moreover, we shall disregard any non-Gaussian contributions in the covariance matrix, although late-time gravitational collapse, as well as the nonlinearity of the 21-cm line as a tracer, may slightly change these results \cite{Munoz:2015eqa}. We then have $\bar{a}^{}_{\ell m} = a^{}_{\ell m} / C_\ell$, and the Fisher matrices for $A_{1M}$ and $g_{2M}$ can be reduced to
\begin{eqnarray}
  F_{1M, 1 M'}^{(A)} &=&
  \frac{\delta_{M,M'}}{6}
  \sum_{\ell_1, \ell_2 = \ell_{\rm min}}^{\ell_{\rm max}} 
  h_{\ell_1 \ell_2 1}^2
  \frac{\left( G_{\ell_1 \ell_1} + G_{\ell_2 \ell_2} \right)^2}{C_{\ell_1} C_{\ell_2}}~, \label{eq:Fish_A1M}
  \\
 F_{2M, 2 M'}^{(g)} &=& \frac{\delta_{M, M'}}{10} 
  \sum_{\ell_1, \ell_2 = \ell_{\rm min}}^{\ell_{\rm max}} h_{\ell_1 \ell_2 2}^2 
\frac{G_{\ell_1 \ell_2}^2}{C_{\ell_1} C_{\ell_2}} ~. \label{eq:Fish_g2M}
\end{eqnarray}
In the following discussions, we analyze the expected $1\sigma$ errors on $A_{1M}$ and $g_{2M}$, computing $\sigma (A_{1M}) = (F_{1M, 1 M}^{(A)})^{-1/2}$ and $\sigma( g_{2M}) = (F_{2M, 2 M}^{(g)})^{-1/2}$. The minimum multipole is taken to be $\ell_{\rm min} = 2$ in this paper, while contaminations of low-$\ell$ data due to residual galactic synchrotron emission could raise $\ell_{\rm min}$ in realistic data analyses. This could diminish the sensitivities; however, the cases with blue-tilted $f(k)$ could be relatively less affected.

Figure~\ref{fig:err} shows the $\ell_{\rm max}$ dependence of $\sigma (A_{1M})$ and $\sigma( g_{2M})$ for $z = 30$ and $\Delta \nu = 0.1 \, {\rm MHz}$. 
As seen in this figure, the sensitivities to $A_{1M}$ and $g_{2M}$ for the FRA get better as $\ell_{\rm max}$ increases up to $\ell_{\rm max} \sim 10^4$, after which the sensitivity plateaus due to the noise becoming dominant. In fact, the sensitivities for the FRA and CVL cases are comparable for $\ell_{\rm max} \lesssim 10^3$. 
The FRA sensitivities exceed the SKA ones by $\sim 4-5$ orders of magnitude since the error bars scale like $t_o^{-1} f_{\rm cover}^{-2} D_{\rm base}^{-2} $; thus, highly accurate measurements are possible with FRA, being the constraints comparable to, or better than, {\it Planck} 2015 \cite{Kim:2013gka,Ade:2015sjc,Ade:2015lrj}. The outperformance of FRA is simply due to big improvements of $f_{\rm cover}$ and $D_{\rm base}$. We here assume that contaminations due to foregrounds are completely subtracted and find the best results, while residual components could get the sensitivities somewhat worse in real experiments. In the two top panels for $\sigma(A_{1M})$, we notice that the sensitivities in the CVL cases drastically increase beyond the pivot scale $\ell_A \simeq 1.3 \times 10^4$ as expected from Fig.~\ref{fig:Cls}. Likewise, in the blue-tilted cases for $\sigma(g_{2M})$ ($f(k) \propto k^{1,2}$), the error bars decline drastically at around $\ell_g \simeq 650$, corresponding with the crossing point in Fig.~\ref{fig:Cls}, as seen in the two bottom panels.

For the CVL cases, the scalings of $\sigma (A_{1M})$ and $\sigma( g_{2M})$ can be estimated analytically in the same manner as Ref.~\cite{Pullen:2007tu}. For the dipolar asymmetry case with $f(k) = (1 - k/k_A)^{\{1,2\}}$, we can approximate $G_{\ell, \ell} + G_{\ell \pm 1, \ell \pm 1} \simeq 2 C_\ell$ (see Fig.~\ref{fig:Cls}), which is valid for $\ell \lesssim \ell_A$. Since the selection rule restricts the range of $\sum_{\ell_2}$ to $\ell_2 = \ell_1 \pm 1$, the diagonal elements of the Fisher matrix \eqref{eq:Fish_A1M} are reduced to
\begin{eqnarray}
 F_{1M, 1M}^{(A)} \simeq \frac{2}{3} h_{\ell_{\rm min}, \ell_{\rm min}+1, 1}^2
  +  \frac{2}{3}
  \sum_{\ell_1 = \ell_{\rm min} + 1}^{\ell_{\max}-1}
  \left( h_{\ell_1, \ell_1 - 1, 1}^2 + h_{\ell_1, \ell_1 + 1, 1}^2  \right)
  + \frac{2}{3} h_{\ell_{\max}, \ell_{\max}-1, 1}^2 ~.
\end{eqnarray}
For $\ell_{\rm max} \gg \ell_{\rm min}$, this yields $\sigma(A_{1M}) \simeq \sqrt{2\pi} / \ell_{\rm max}$, matching up with the purple lines of the top two panels in Fig.~\ref{fig:err} for $\ell < \ell_A$. Likewise, for $f(k) = 1$, using the high-$\ell$ approximation $G_{\ell, \ell} \simeq G_{\ell, \ell \pm 2} \simeq C_{\ell}$, and the selection rules of $h_{\ell_1 \ell_2 2}$, we can simplify the Fisher matrix for $g_{2M}$ \eqref{eq:Fish_g2M} to
\begin{eqnarray}
 F_{2M, 2 M}^{(g)} &\simeq& \frac{1}{10} \left(
  h_{\ell_{\rm min} \ell_{\rm min} 2}^2 + h_{\ell_{\rm min}, \ell_{\rm min}+2, 2}^2
  + h_{\ell_{\rm min}+1, \ell_{\rm min}+1, 2}^2 + h_{\ell_{\rm min}+1, \ell_{\rm min}+3, 2}^2
  \right) \nonumber \\ 
&&  + \frac{1}{10} \sum_{\ell_1 = \ell_{\rm min}+2}^{\ell_{\rm max}-2} \left(
  h_{\ell_1, \ell_1-2, 2}^2 + h_{\ell_1, \ell_1, 2}^2 + h_{\ell_1, \ell_1+2, 2}^2
  \right) \nonumber \\
  &&+ \frac{1}{10} \left(
  h_{\ell_{\rm max}-1, \ell_{\rm max}-3, 2}^2 + h_{\ell_{\rm max}-1, \ell_{\rm max}-1, 2}^2
  + h_{\ell_{\rm max}, \ell_{\rm max}-2, 2}^2 + h_{\ell_{\rm max} \ell_{\rm max} 2}^2
  \right) ~,
\end{eqnarray}
leading to $\sigma(g_{2M}) \simeq \sqrt{8\pi} / \ell_{\rm max}$ for $\ell_{\rm max}\gg \ell_{\rm min}$. This also recovers the purple line of the center panel in Fig.~\ref{fig:err} for high $\ell_{\rm max}$, validating our numerical treatments. The identical estimates are obtained in a CMB CVL-level measurement \cite{Pullen:2007tu}; however, the 21-cm analysis can, in principle, go to higher $\ell_{\rm max}$ and further reduce both $\sigma(A_{1M})$ and $\sigma(g_{2M})$. If taking into account an instrumental noise or the scale dependence of $f(k)$, the scaling of the error bars deviates from $\propto \ell_{\rm max}^{-1}$, as seen in Fig.~\ref{fig:err}.

One of the advantages of using the 21-cm line is being able to use tomography, i.e., to co-add information from different redshift slices. Given a certain frequency range (from $\nu_1$ to $\nu_2$), we can add all bandwidths $\Delta\nu$ within that range. More precisely, we can do so if the information in each redshift slice is uncorrelated. The correlation length depends on $z$ and $\ell$, and it is at most $\sim 0.5 \, {\rm MHz}$ for $20 \leq z \leq 50$ and $\ell \geq \ell_{\rm min} = 2$ \cite{Munoz:2015eqa}. In our tomographic analysis, we take only the larger bandwidth $\Delta \nu = 1.0 \, {\rm MHz}$ to keep uncorrelated slices. In this case, we can write the final Fisher matrix as a sum over all central frequencies $\nu_i$ and approximate that by an integral over redshifts as
\be
\sum_{\nu_i} F(\nu_i) \approx \int_{\nu_1}^{\nu_2} \dfrac{d\nu}{\Delta \nu}F(\nu) ={ \int_{z_2}^{z_1} } \dfrac{dz}{(1+z)^2} \dfrac{\nu_0}{\Delta\nu}F(z),  
\ee
where $\nu_0=1420 \, {\rm MHz}$ is the rest frequency of the 21-cm transition, and $F(z)$ is the Fisher matrix at redshift $z$. 
The redshift evolution of $\sigma(A_{1M})$ and $\sigma(g_{2M})$ at $\ell_{\rm max} = {\rm Min}[10^5, \ell_{\rm cover}]$ for $\Delta \nu = 1.0 \, {\rm MHz}$ is described in Fig.~\ref{fig:err_dip-quad_z}, showing a weak dependence on $z$ for all CVL constraints and an increase of noise at higher redshifts. We find there that the single-slice FRA sensitivities are comparable to the sensitivities in a noiseless CVL-level measurement of CMB temperature and E-mode polarization (T+E) anisotropies. 

The results obtained by integrating all redshift slices between $z = 20$ and $z = 50$ are summarized in Tables~\ref{tab:results_A1M} ($A_{1M}$) and \ref{tab:results_g2M} ($g_{2M}$). There we find that, owing to tomography, the sensitivities to both $A_{1M}$ and $g_{2M}$ are improved by a factor of $\sim 5$ compared to single-slice results. Due to this, the FRA sensitivities can surpass the sensitivities achieved in a noiseless CMB measurement, even in a red-tilted $f(k) \propto k^{-1}$ case of $g_{2M}$. A noiseless CVL 21-cm measurement leads to more drastic improvements of sensitivity, surpassing the CVL CMB constraints (except for the $f(k) \propto k^{-2}$ case of $g_{2M}$), due to the great quantity of available $\ell$ modes and multi-$z$ information added by tomography.

\begin{table}[h]
{\small
\begin{tabular}{|c||c|c|c||c|c|}
\hline
$f(k)$ & CVL 21 cm & SKA & FRA & CVL CMB T & CVL CMB T+E \\ \hline\hline
$ 1 - k/k_A$ & $8.0 \times 10^{-7}$ ($5.1 \times 10^{-6}$) & $13$ ($62$) & $6.4 \times 10^{-4}$ ($3.3 \times 10^{-3}$) & $1.3 \times 10^{-3}$ & $9.3 \times 10^{-4}$ \\
$(1 - k/k_A)^2$ & $1.4 \times 10^{-7}$ ($8.7 \times 10^{-7}$) & $14$ ($64$) & $6.9 \times 10^{-4}$ ($3.6 \times 10^{-3}$) & $1.5 \times 10^{-3}$ & $1.0 \times 10^{-3}$ \\ 
\hline
\end{tabular}
}
\caption{Expected $1\sigma$ errors $\sigma(A_{1M})$ summing from $\ell_{\rm min}=2$ to $\ell_{\rm max} = {\rm Min}[10^5, \ell_{\rm cover}]$, integrating all redshift slices between $z = 20$ and $z = 50$, with $\Delta \nu = 1.0 \, {\rm MHz}$ and $k_A = 1 \ {\rm Mpc^{-1}}$. For comparison, we show the values with a single slice at $z = 30$ in parentheses. The first three columns present the results obtained in the measurements of 21-cm fluctuations, while the last two columns show the results we obtain for a noiseless CVL-level CMB measurement up to $\ell = 2000$, by analyzing temperature anisotropy only (T) and temperature and E-mode polarization jointly (T+E).}
\label{tab:results_A1M}
\end{table}

\begin{table}[h]
{\small
\begin{tabular}{|c||c|c|c||c|c|}
\hline
$f(k)$ & CVL 21 cm & SKA & FRA & CVL CMB T & CVL CMB T+E \\ \hline\hline
$(k/k_g)^2$ & $5.0 \times 10^{-10}$ ($3.2 \times 10^{-9}$) & $4.2$ ($22$) & $ 1.4 \times 10^{-5}$ ($6.6 \times  10^{-5}$) & $5.5 \times 10^{-4} $ & $3.2 \times 10^{-4}$ \\
$(k/k_g)^1$ & $6.7 \times 10^{-8}$ ($4.3 \times 10^{-7}$) & $20$ (95) & $3.6 \times 10^{-4}$ ($1.7 \times 10^{-3}$) & $1.5 \times 10^{-3}$ & $8.3 \times 10^{-4}$ \\
$1$ & $7.9 \times 10^{-6}$ ($5.0 \times 10^{-5}$) & $33$ ($150$) & $1.3 \times 10^{-3}$ ($ 6.3 \times 10^{-3}$) & $3.4 \times 10^{-3}$ & $1.9 \times 10^{-3}$ \\
$(k/k_g)^{-1}$ & $3.8 \times 10^{-4}$ ($2.4 \times 10^{-3} $) & $17$ ($78$) & $1.3 \times 10^{-3}$ ($7.2 \times 10^{-3}$) & $4.3 \times 10^{-3}$ & $2.1 \times 10^{-3}$ \\
$(k/k_g)^{-2}$ & $3.2 \times 10^{-4} $ ($2.0 \times 10^{-3}$) & $3.4$ ($16$) & $4.2 \times 10^{-4}$ ($2.4 \times 10^{-3}$) & $6.1 \times 10^{-5}$ & $3.7 \times 10^{-5}$ \\
\hline
\end{tabular}
}
\caption{
Expected $1\sigma$ errors $\sigma(g_{2M})$ with $k_g = 0.05 \ {\rm Mpc^{-1}}$ but otherwise the same parameters as Table \ref{tab:results_A1M}.
}
\label{tab:results_g2M}
\end{table}

\section{Conclusions}
\label{sec:conclusions}
In this paper, we studied the constraining power of 21-cm fluctuations for detecting primordial rotational asymmetries, in particular the dipolar and quadrupolar asymmetries. In principle, 21-cm fluctuations contain information on density fluctuations up to very small scales, far beyond scales accessible to the CMB. Moreover, these fluctuations allow for multi-redshift analyses, enhancing sensitivity compared to the CMB, which is a single 2D screen and is damped at much lower $\ell$.

We confirmed that dipolar and quadrupolar rotational asymmetries create nonvanishing off-diagonal correlations in the angular power spectrum of 21-cm fluctuations. By means of a Fisher matrix analysis, we estimated the minimum value of the magnitude of those asymmetries detectable for several types of scale dependencies. We considered three different experimental setups to forecast these constraints. The planned SKA radio survey and a somewhat futuristic radio array, which we denoted as FRA. Finally, in order to understand what will be, in principle, possible to measure with 21-cm experiments, we also investigated constraints coming from a cosmic variance limited (CVL) 21-cm survey.

A comprehensive analysis demonstrated that, owing to the large number of $\ell$ modes available, both the FRA and CVL surveys could improve upon CMB constraints \cite{Kim:2013gka,Ade:2015lrj,Ade:2015sjc} on a wide range of the dipolar and quadrupolar asymmetries.
The possibility to have tomographic analyses using several redshift slices provides additional information, helping to achieve a sensitivity better than an ideal cosmic variance limited CMB survey for these asymmetry parameters. We found that the SKA could provide some constraining power for asymmetry parameters, even if not competitive with current limits coming from CMB maps; 21-cm measurements would in any case be a useful check, as they will have different systematics and come from an entirely different observable.
As for a futuristic radio array, we found that the minimum amplitude of dipolar modulation measurable was $A_{1M} \lesssim 10^{-3}$ for FRA and $A_{1M} \lesssim 10^{-6}$ for CVL, compared to $\sim 10^{-3}$ for an ideal CMB experiment.
As for quadrupolar models, our results show that the FRA could also constrain the amplitude $g_{2M}$ to be $\lesssim 10^{-3} - 10^{-5}$ (depending on the model), which could be further improved to $\lesssim 10^{-4} - 10^{-10}$ by a CVL survey, compared to $\lesssim 10^{-3} - 10^{-5}$ obtainable with an ideal T+E CMB measurement. 

Moreover, 21-cm surveys provide independent probes of broken rotational invariance and as such would help disentangle potential biases present in previous CMB experiments.

In this work we focused on autocorrelations of the 21-cm fluctuations; however, it is expected that cross-correlations between 21-cm fluctuations and other observables, such as CMB anisotropies or galaxy number counts, could provide additional information on cosmological rotational asymmetries. Moreover, there may be nontrivial rotationally asymmetric signatures (violating the usual triangular conditions in harmonic space) on higher-order correlations (such as the bispectrum and the trispectrum) of 21-cm fluctuations as well as of the CMB  \cite{Shiraishi:2011ph,Bartolo:2011ee}. These prospects are left to be studied in a future work.


\acknowledgements


MS was supported in part by a Grant-in-Aid for JSPS
Research under Grant No.~27-10917, and in part by World Premier
International Research Center Initiative (WPI Initiative), MEXT,
Japan.
JBM, MK and AR were supported at JHU by NSF Grant No.\ 0244990, NASA
NNX15AB18G, the John Templeton Foundation, and the Simons
Foundation.


\appendix

\section{Flat-sky limit}
\label{appen:flat-sky}
In the flat-sky approximation, the temperature fluctuations in Fourier space are given by
\be
\delta T(\bfell) = \int \dfrac{d^3k}{(2\pi)^3} e^{i r k_{||}} \tilde W(k_{||})
{\delta T({\bf k})}
(2\pi)^2 \tilde \delta_D^{(2)}(r \mathbf k_{\perp} - \bfell),
\ee
where $\tilde W(k_{||})$ is the window function transformed to Fourier space, and we have defined 
\be
\tilde \delta_D^{(2)}{(\bfell)} \equiv \dfrac{1}{(2\pi)^2} \int_A \dfrac{d^2 x}{r^2} e^{i \mathbf x\cdot \bfell/r},
\ee
which has a value of $\tilde \delta_D^{(2)}(0)=\fsky/\pi$ at the origin, and its convolution with itself 
returns the same function and has a characteristic width $\delta\ell\sim 1/\sqrt{\fsky}$. For us to find the covariance matrix, we would need to correlate
\be
\VEV{\delta T(\bfell_1)\delta T(\bfell_2)} = \int\dfrac{d^2k_{\perp}}{(2\pi)^2} (2\pi)^2 \tilde \delta_D^{(2)}(r \mathbf k_{\perp} - \bfell_1) (2\pi)^2 \tilde \delta_D^{(2)}({-r} \mathbf k_{\perp} - \bfell_2) f(\mathbf k_{\perp}),
\ee
with $f(\mathbf k_{\perp}) = \int dk_{||}/(2\pi) |\tilde W(k_{||})|^2 P_{\delta T}{({\bf k})}$. We can take $f(\mathbf k_{\perp}) \approx {f(\bfell_1 / r)}$, so then 
\be
G_{\ell_1,\ell_1+\Delta \ell} = G_{\ell_1,\ell_1} \dfrac{\pi}{\fsky} \times \int\dfrac{d^2x}{(2\pi)^2r^2} e^{i \mathbf x \cdot \Delta \bfell/r},
\label{eq:Gellint}
\ee
where $r$ is the distance to the surface of observation, $R$ is the radius of the piece of sky observed, and then $f_{\rm sky} = R^2/(4 r^2)$.

We can calculate the integral in Eq.~\eqref{eq:Gellint} to be
\be
\int\dfrac{d^2x}{(2\pi)^2r^2} e^{i \mathbf x \cdot \Delta \bfell/r} = \dfrac{1}{(2\pi)^2 r^2} \int_0^R dx x (2\pi) J_0(x \Delta\ell/r)= \dfrac{R}{(2\pi) r} \dfrac{J_1(\Delta\ell R/r)}{\Delta\ell},
\ee
this quantity is, to second order in $\Delta \ell$, given by
\be
\int\dfrac{d^2x}{(2\pi)^2r^2} e^{i \mathbf x \cdot \Delta \bfell/r} = \dfrac{\fsky}{\pi} \left [1-\dfrac{\fsky}{2} (\Delta\ell)^2\right],
\ee
so finally,
\be
G_{\ell_1,\ell_1+\Delta \ell} = G_{\ell_1,\ell_1} \times \left[1-\dfrac{\fsky}{2} (\Delta\ell)^2 \right] ,
\ee
so, in the flat-sky limit ($\fsky\ll 1$), we can just take $G_{\ell_1,\ell_1+\Delta \ell} = G_{\ell_1,\ell_1}$.

\bibliography{paper}


\end{document}